# Geodetic Research on Deception Island and its Environment (South Shetland Islands, Bransfield Sea and Antarctic Peninsula) During Spanish Antarctic Campaigns (1987–2007)

M. Berrocoso, A. Fernández-Ros, M.E. Ramírez, J.M. Salamanca, C. Torrecillas,
A. Pérez-Peña, R. Páez, A. García-García, Y. Jiménez-Teja, F. García-García,
R. Soto, J. Gárate, J. Martín-Davila, A. Sánchez-Alzola, A. de Gil,
J.A. Fernández-Prada and B. Jigena

---

M. Berrocoso
Laboratorio de Astronomía, Geodesia y Cartografía. Departamento de Matemáticas.
Facultad de Ciencias. Campus de Puerto Real. Universidad de
Cádiz. 11510 Puerto Real (Cádiz-Andalucía). España, e-mail: manuel.berrocoso@uca.es

A. Fernández-Ros
Laboratorio de Astronomía, Geodesia y Cartografía. Departamento de Matemáticas.
Facultad de Ciencias. Campus de Puerto Real. Universidad de Cádiz. 11510 Puerto Real
(Cádiz-Andalucía). España

M.E. Ram´ırez
Laboratorio de Astronomía, Geodesia y Cartografía. Departamento de Matemáticas.
Facultad de Ciencias. Campus de Puerto Real. Universidad de
Cádiz. 11510 Puerto Real (Cádiz-Andalucía). España

J.M. Salamanca
Laboratorio de Astronomía, Geodesia y Cartografía. Departamento de Matemáticas.
Facultad de Ciencias. Campus de Puerto Real. Universidad de
Cádiz. 11510 Puerto Real (Cádiz-Andalucía). España

C. Torrecillas
Laboratorio de Astronomía, Geodesia y Cartografía. Departamento de Matemáticas.
Facultad de Ciencias. Campus de Puerto Real. Universidad de
Cádiz. 11510 Puerto Real (Cádiz-Andalucía). España.

A. Pérez-Peña
Laboratorio de Astronomía, Geodesia y Cartografía. Departamento de Matemáticas.
Facultad de Ciencias. Campus de Puerto Real. Universidad de Cádiz. 11510 Puerto Real
(Cádiz-Andalucía). España

R. Páez
Laboratorio de Astronomía, Geodesia y Cartografía. Departamento de Matemáticas.
Facultad de Ciencias. Campus de Puerto Real. Universidad de Cádiz. 11510 Puerto Real
(Cádiz-Andalucía). España

A. Garc´ıa-Garc´ıa
Departamento de Volcanolog´ıa. Museo Nacional de Ciencias Naturales. Consejo Superior
de Investigaciones Científicas. C/ José Gutiérrez Abascal, 2. Madrid

Y. Jiménez-Teja
Laboratorio de Astronomía, Geodesia y Cartografía. Departamento de Matemáticas.
Facultad de Ciencias. Campus de Puerto Real. Universidad de Cádiz. 11510 Puerto Real
(Cádiz-Andalucía). España

**Abstract** Since 1987, Spain has been continuously developing several scientific projects, mainly based on Earth Sciences, in Geodesy, Geochemistry, Geology or Volcanology. The need of a geodetic reference frame when doing hydrographic and topographic mapping meant the organization of the earlier campaigns with the main goals of updating the existing cartography and of making new maps of the area. During this period of time, new techniques arose in Space Geodesy improving the classical methodology and making possible its applications to other different fields such as tectonic or volcanism. Spanish Antarctic Geodetic activities from the 1987–1988 to 2006–2007 campaigns are described as well as a geodetic and a levelling network are presented. The first network, RGAE, was designed and established to define a reference frame in the region formed by the South Shetlands Islands, the Bransfield Sea and the Antarctic Peninsula whereas the second one, REGID, was planned to control the volcanic activity in Deception Island. Finally, the horizontal and vertical deformation models are described too, as well as the strategy which has been followed when computing an experimental geoid.

___________________

F. García-García

Escuela Superior de Ingeniería Cartográfica y Geodésica. Universidad Politécnica de Valencia

R. Soto

Servicio de Satélites. Sección de Geofísica. Real Instituto y Observatorio de la Armada. San Fernando (Cádiz, España)

J. Gárate

Servicio de Satélites. Sección de Geofísica. Real Instituto y Observatorio de la Armada. San Fernando (Cádiz, España)

J. Martn-Davila

Servicio de Satélites. Sección de Geofísica. Real Instituto y Observatorio de la Armada. San Fernando (Cádiz, España)

A. Sánchez-Alzola

Laboratorio de Astronomía, Geodesia y Cartografía. Departamento de Matemáticas. Facultad de Ciencias. Campus de Puerto Real. Universidad de Cádiz. 11510 Puerto Real (Cádiz-Andalucía). España

A. de Gil

Laboratorio de Astronomía, Geodesia y Cartografía. Departamento de Matemáticas. Facultad de Ciencias. Campus de Puerto Real. Universidad de Cádiz. 11510 Puerto Real (Cádiz-Andalucía). España

J.A. Fernández-Prada

Laboratorio de Astronomía, Geodesia y Cartografía. Departamento de Matemáticas. Facultad de Ciencias. Campus de Puerto Real. Universidad de Cádiz. 11510 Puerto Real (Cádiz-Andalucía). España

B. Jigena

Laboratorio de Astronomía, Geodesia y Cartografía. Departamento de Matemáticas. Facultad de Ciencias. Campus de Puerto Real. Universidad de Cádiz. 11510 Puerto Real (Cádiz-Andalucía). España

# 1 Geodetic Reference Frame for South Shetland Island, Bransfield Sea and Antarctic Peninsula

Meteorological, oceanographic, geophysical, geodynamic, biological, glaciology, etc., all require reference frame that establishes a precise time and space of the data. Due to its geographical isolation, realising accurate reference frames in Antarctica has been a large challenge.

The beginning of the Global Positioning System (GPS) and its functionality on April 27, 1985, constituted an important improvement in the productivity and the obtained precision in relation to other spatial positioning systems. It also meant a reduction in the cost of the equipments and field programmes. GPS is based in the interferometric principle of simultaneous observations of NAVSTAR constellation, which provides a reference system called WGS-84 ellipsoid (NIMA 2000), and obtain a relative positioning between the stations with precisions of 1 ppm. To obtain absolute precisions, a first level network formed by geodesic stations situated near VLBI stations is established. Later, a relative positioning of the stations is realized in respect of the stations which belong to that network. This positioning gives absolute coordinates using a later adjustment.

The use of the reference ellipsoid WGS-84, implies the unify of the local reference systems established in different zones of the Earth; In the other hand, the ellipsoid coordinates of the stations are obtained, and later, the values of vertical deviation and geoid undulation are calculated. These values allow obtaining the astronomical coordinates and their respective ortometric altitudes for each station.

In the Antarctic environment (Berrocoso 1997), the lack of a local datum, obtained by direct observation, and therefore, a local ellipsoid, implied that the adoption of global ellipsoid WGS-84 presented some disadvantages. In the Antarctic area we can not obtain transformation parameters between the global and the local ellipsoid for use it to secondary stations.

In 1989–1990 campaign, a fundamental geodetic station was constructed in Juan Carlos I Antarctic Base (Livingston Island), by means of monitoring TRANSIT satellites. The absolute coordinates of this station were calculated by applying Doppler precise point positioning, referred to the WGS-72 Geodetic System (NIMA 2000). These coordinates were transformed to WGS-84 to calculate the coordinates of the other stations by means of relative GPS positioning. During 1989–1990 campaign, a prolongation of the South American Geodetic Network was carried out. A link between Tierra del Fuego and South Shetland Island was made using GPS measurements, where a point constructed in Rio Grande Astronomical Station was elected as the principal point of the link.

The validity of this work was extended and improved during the austral summer of 1991–92, with the accomplishment of the international campaign SCAR'92 (*Scientific Commission Antarctic Researches*). This campaign allowed the establishment of the final Global Geodetic Reference Frame in Antarctic in Geodetic Level A (Seeber 2003) whose station coordinates was provided with accuracy of the order of centimetres in its absolute coordinates.

## 1.1 The Spanish Antarctic Geodetic Network

The Spanish Antarctic Geodetic Network, RGAE network consists of several stations around the South Shetland Island, the Bransfield Sea and the Antarctic Peninsula. These stations have precise absolute coordinates and they are referred to WGS84 ellipsoid in the ITRF reference frame corresponding. To the initial objective of being the regional geodetic reference frame (due to the accuracies reached in the relative positioning), the objective of establishment the geodynamical frame was added. The GPS surveying of this network will provide the tectonic behaviour of the region.

### 1.1.1 Design and Evolution of the Geodetic Network RGAE

The design of the geodetic network RGAE was planned taking into account the available equipments and the special conditions of the area. Other important aspects in the location of the stations of the RGAE geodetic network due to the isolated characteristic of the environment, was the logistic, the accessibility of the station, the electric management of the instruments, a sky free from obstacles, the existence of problems in the signal and the safety of the researchers. These considerations mint that this geodetic network, the most extensive in the Antarctic area of Level B, has a specific design with accuracy similar to other networks of classical designs (Seeber 2003). It is important to emphasize that the stations were built on geodetic pillars, with a fix screw to insure the precise placement of the GPS antenna.

Due to the lack of satellites in the GPS constellation during the first campaigns, the observations were planned depending on the geometric configuration of the satellites: the choice of the suitable satellites ensured simultaneous visibility between the points, and the storage capacity of the receiver.

From 1995–1996 campaign, GPS receiver had more storage capacity and lower power requirements. These aspects allowed a greater flexibility in the planning of the GPS surveys (sessions of continue 24 h in all of the campaign). In the latest campaigns, 1 Hz sampling rate and a 0° elevation mask are available and some of the stations and continuously recording data. It is important to emphasize that the surveys were made with dual frequency geodetic receivers in order to remove the ionospheric effect from the observations. Figure 1 shows the site locations.

In the Antarctic campaign 1987–1988 the TRANSIT satellite observations were recorded by JMR-1 receivers, allowing the obtaining of absolute coordinates (ellipsoid WGS-72) using the Doppler Punctual Positioning Method (accuracies of 3–10 m). During the 1988–1989 campaign every station was provided by absolute geocentric coordinates, obtaining an accuracy of 5 m approximately. In the 1989–1990 campaign, RIOG point station was set as the fixed station in the network adjustment, obtaining an accuracy of 1–2 m. During the 1990–1991 campaign, RIOG (Argentina) and PUAR (Chile) were fixed in the processing to obtain the absolute coordinates of the rest of the stations in the RGAE network.

The validity of this approach was confirmed during the austral summer of 1991–1992 with the realization of the SCAR'92 international campaign, whose main objectives were the establishment of an Antarctic geodetic network covering the whole area, and connected to the IER world network, an extension of the Australian, the New-Zealander, the South-African and the South-American network. During this campaign the final Antarctic polyhedron for a global geodetic reference frame of A Level were established. Its stations covered the Polar Cap and they were provided by absolute coordinates with precisions of the order of one centimetre. Although USHU and BARG were observed, they were not consider in the final adjustment since they were too close compared to the rest of the stations in the network. The calculus and the adjustment of the network were realized by Canberra University and the resulting absolute coordinates obtained were referred to the ITRF 1992.02 frame.

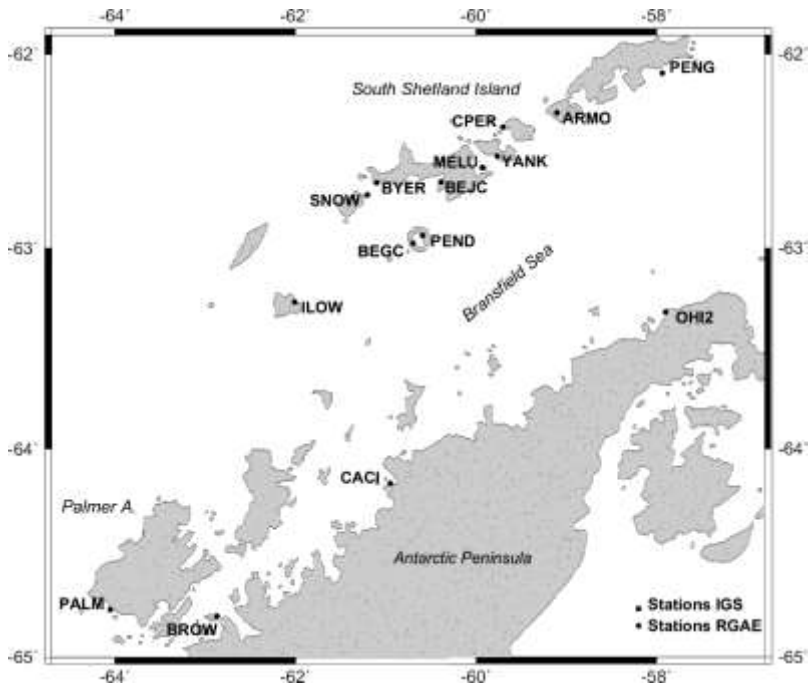

**Fig. 1** Distribution of the stations of the REGAE network

A volcanic event crisis occurring on Deception Island from December 1991 to January 1992 motivated the monumentation of another station in the island, BEGC at the Spanish Base Gabriel de Castilla, which would be included in the RGAE network. In 1999–2000 campaign OHIG mark as the reference point. During 2003–2004 campaign, the observations were realized with a 1 Hz sampling rate and a 0° elevation mask (Table 1).

At present, RGAE geodetic network consists of the following stations: BEJC and BYER (Livingston Island), BEGC (Deception Island), ILOW (Low Island), SNOE (Snow Island), LUNA (Half Moon Island), PRAR and YANK (Greenwich

**Table 1** Summary of the occupation for GPS campaigns

| ID | Situation | 87–88 | 88–89 | 89–90 | 90–91 | 91–92 | 99–00 | 03–04 | 06–07 |
|---|---|---|---|---|---|---|---|---|---|
| BEJC | Livingston I. | X | – | X | X | X | X | X | X |
| BARG | Deception I. | X | – | X | X | X | X | X | X |
| PUPO | Livingston I. | – | X | X | – | – | – | – | – |
| FUMA | Deception I. | – | X | – | X | X | X | X | X |
| PRAT | Antarctic P. | – | X | – | – | – | – | – | X |
| BREJ | King George I. | – | – | X | – | – | – | – | – |
| PALM | Anvers I. | – | – | X | – | – | – | – | – |
| RIOG | Argentina | – | – | X | X | X | – | – | – |
| USHU | Argentina | – | – | X | X | X | – | – | – |
| PEND | Deception I. | – | – | – | X | X | X | X | X |
| BALL | Deception I. | – | – | – | X | X | X | X | X |
| PUAR | Chile | – | – | – | X | – | – | – | – |
| OHIG | Antarctic P. | – | – | – | X | – | – | – | – |
| BEGC | Deception I. | – | – | – | – | – | X | X | X |
| CACI | Antarctic P. | – | – | – | – | – | – | X | X |
| BROW | Antarctic P. | – | – | – | X | – | – | – | X |
| ILOW | Low I. | – | – | – | – | – | – | – | X |
| SNOW | Snow I. | – | – | – | – | – | – | – | X |
| BYER | Livingston I. | – | – | – | – | – | – | – | X |
| ARMO | Nelson I. | – | – | – | – | – | – | – | X |
| MELU | Half Moon I. | – | – | – | – | – | – | – | X |
| YANK | Greenwich I. | – | – | – | – | – | – | – | X |
| PENG | Penguin I. | – | – | – | – | – | – | – | X |
| CPER | Robert I. | – | – | – | – | – | – | – | X |

Island), CPER (Robert Island), ARMO (Nelson Island), PING (Penguin Island), CACI (Cierva Cove, Antarctic Peninsula), ALBR (Antarctic Peninsula). Other stations previously considered are now administered for other countries, as OHIG and PALM stations, at the O'Higgins Chilean Base and at the American Antarctic Base Palmer on Anvers Island, respectively (Fig. 1).

From 1991–1992 campaign, several episodic campaigns (organized for SCAR) have been realized. With the goal of contributing to the redefinition of the Antarctic reference frame, these campaigns were carried out from January 20th to February 10th each year. The configurations of the observations are: 24 h sessions, 0° elevation mask and 1 s sampling rate. Since 1996, GPS observations from BEJC and BEGC are included in episodic campaigns: EPOCH'96, EPOCH'02, EPOCH'03, EPOCH'04, EPOCH'05, EPOCH'06 and EPOCH'07.

### 1.1.2 Data Processing and Adjustment of the Network

The data coming from the first campaigns were processed with software TRIMMBP (Trimble 1991). And from campaign 1991–1992 all the data have been reprocess with Bernese v4.2 scientific software (Hugentobler et al. 2001), developed by Astronomical Institute at the University of Berne in Switzerland. It combines specialized surveying knowledge with advanced software techniques. Standard procedures were applied by using IGS products. Final IGS orbits and Earth Rotation Parameters were used according to the proceedings of the Fourth Analysis Centres Workshop in Graz, Austria, September 2003.

Stations coordinate are estimated considering 24 h sessions, 10° elevation masks and a 30 s sampling rate (Hugentobler et al. 2006). Since the software is based on relative positioning, the daily set of baselines were combined in a network adjustment which involves every station for each day. During the parameter estimation process, carrier-phase double-difference data is used in an ionospheric delay free mode.

Tropospheric errors are dealt with by using a combination of the a priori Saastamoinen model and Neill mapping functions. Tropospheric parameters are hourly estimated, ambiguities are dealt for each baseline independently, using the ionosphere free observable. The QIF algorithm is used to solve ambiguities (Mervart 1995). QIF ambiguity resolution requires L1 and L2 to be processed in parallel rather than processing the ionosphere free L3 linear combination. An a priori ionosphere model, ocean tide loading displacement corrections from Onsala Observatory (*www.oso.chalmers.se/ loading/*) and tropospheric parameters are introduced.

Finally, we calculate the station coordinates and normal equations for the daily solutions. ADDNEQ was used to combine normal equations. Two strategies have been applied for the solution adjustment at each epoch: free network adjustment and heavily constrained. The first solution was used to analyse the quality of the coordinates by comparing the IGS fiducial station coordinates in both reference system. Constrained solutions were processed fixing IGS (*International Geodesy Station*) stations to the ITRF2000 reference frame (Altamimi et al. 2002) at the mean epoch of each campaign.

The coordinates for the stations in RGAE network referred to ITRF2000 are listed in Table 2. The stations of network RGAE that are not indicated in this table they have not been processed yet, since it has been surveyed during the last Antarctic campaign for the first time.

**Table 2** Coordinates of stations RGAE network (epoch 2006.5)

| Station | X (mts.) | $\sigma_X$ | Y (mts.) | $\sigma_Y$ | Z (mts.) | $\sigma_Z$ |
|---|---|---|---|---|---|---|
| BEGC | 1423027.717 | 0.004 | −2533143.911 | 0.004 | −5658977.619 | 0.008 |
| BEJC | 1451089.570 | 0.004 | −2553226.315 | 0.008 | −5642854.567 | 0.017 |
| CACI | 1353462.767 | 0.003 | −2437400.371 | 0.006 | −571735.142 | 0.016 |
| FUMA | 1421994.784 | 0.005 | −2535674.066 | 0.009 | −2535674.066 | 0.018 |
| PEND | 1428697.780 | 0.007 | −2534792.711 | 0.011 | −5656765.155 | 0.023 |
| PRAT | 1492479.572 | 0.085 | −2550306.205 | 0.076 | −5633422.474 | 0.080 |
| BROW | 1237367.299 | 0.085 | −2414926.681 | 0.076 | −5752840.021 | 0.080 |

## 1.2 Application of the RGAE Geodetic Network for the Obtention of a Tectonic Model of the Area

The RGAE geodetic network constitutes the geodetic reference frame for the spatial positioning in the South Shetland Islands, the Bransfield Sea and the Antarctic Peninsula. In fact, the maps of the Spanish Antarctic Cartography about Livingston Island and Deception Island were realized according to the stations of the network. Several organizations participated in the cartography elaboration: The Geographic Service of the Spanish Army, the Autonomous University of Madrid in collaboration with the British Antarctic Survey, the Hydrographical Institute of the Spanish Army and the Argentinean Hydrographical Institute.

One in the most interesting application of the RGAE network is the determination of the superficial deformation models for the tectonic activity of the area, which is a very complex tectonic zone where several tectonic plates converge: The South American Plate, the Antarctic Plate and three minor plates, Scotia micro-plate, Phoenix micro-plate and South Shetland micro-plate (Baraldo 1999).

The Scotia micro-plate is characterized by the presence of vertical faults in its northern and southern boundaries, which separate it from the subduction generated in the West of the South American coast. The East boundary is defined by the presence of a back-arch ridge that separates it from the micro-plate of the South Sandwich Islands. This micro-plate subducts under the South American Plate to the East. The North limit of the Scotia Arch presents a relative movement to the East with respect to the South American.

The South boundary has vertical faults but has some differences to the North boundary. The South boundary is characterized by the presence of an extension basin (pull-apart type) like the Protector, Dove or Scan basins (Galindo-Zald´ıvar et al. 2006). In the Southwest, the micro-plate borders to the vertical Shackleton Fracture Area, this moves apart from the Phoenix and the South Shetland micro-plate. In this area there are several tectonic features: The Drake Ridge, the Back-arch basin of the Bransfield Strait and the South Shetland Islands.

The Phoenix micro plate borders the West by the Drake Ridge at North, and in the East on the subduction area of the South Shetland; in the North by the Shackleton Fracture Zone and in the South by the Hero Fracture Zone (Fig. 2). This micro-plate presents a subduction process under the South Shetland basin and the Bransfield Strait. Several studies reveal that the subduction process in the West of the Hero Fracture Zone is no longer active, and thus belongs to the Antarctic Plate.

On the other hand, the Bransfield Central Basin presents a tectonic configuration related to an active subdution. It has an extension area that produces an extension rift with an axis on the NE-SW direction. The central area is the most active of the basin and it borders on the Deception Island at West and Bridgeman Island at East. Two asymmetrical edges limit on the South Shetland Island at North and, on the Antarctic Peninsula at South. It is the 60 km on width, 230 km long and 1950 m deep. The limit of the South Shetland Islands is more abrupt with maximum slope of 25–30°) and it is about 10 km long. (Canals et al. 1997a, González-Ferrán 1985, González-Ferrán 1991).

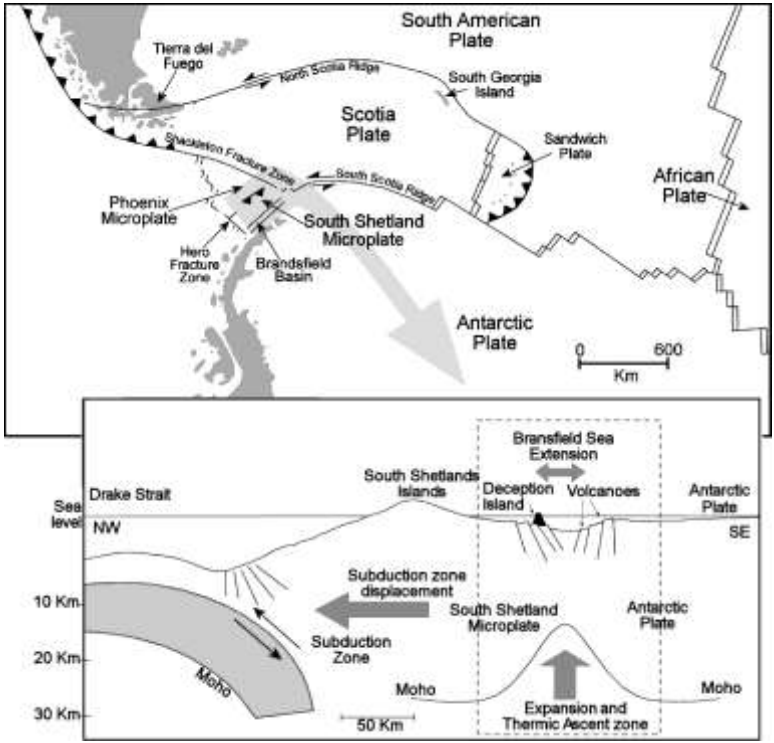

**Fig. 2** Tectonic setting of Deception Island, according to the profile of the Hero fracture

To obtain the superficial deformation models related to the tectonic activity of the area, several stations were resurveyed. To determine the limits of the active subduction area, several new stations were built during the last campaign 2006–2007. To evaluate the tectonic displacement for South Shetland Islands, Bransfield Sea and Antarctic Peninsula, GPS data from 2000/2001 to 2005/2006 campaigns were processed according to the methodology previously describe. OHI2 IGS station was set as the fixed site in the processing.

Figure 3 shows the mean displacement rates for the stations included in the processing. The absolute horizontal velocities are estimated to be at the level of 15 mm./year, except in the CACI and BEJC stations. Few data of CACI station are available since it was surveyed during just two campaigns. The estimated BEJC velocity seems to be anomalous and must be studied in depth when data from the present campaign are be processed.

In addition, Fig. 4 illustrate the displacement rates from 1999/2000 to 2001/2002 campaign, 2001–2002 to 2002–2003 campaign, 2002–2003 to 2003–2004 campaign and 2003–2004 to 2004–2005 campaign respectively. This figure shows an attenuation in the module of the deformation on Deception Island after the seismovolcanic crisis happened in 1998–1999 campaign. It can be watched that the displacement direction vector from BEGC station trends to be equal to the Antarctic Plate, which is given by OHI2 and PALM stations.

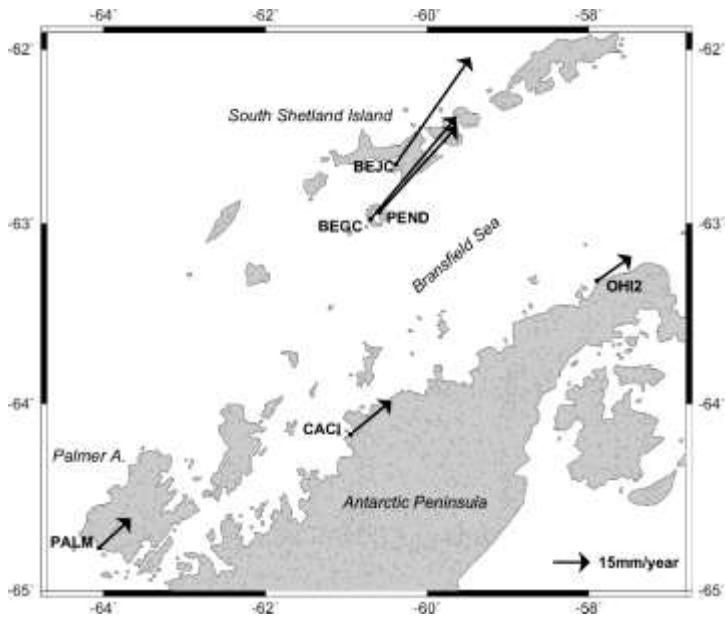

**Fig. 3** Mean displacement rate of some of the stations in RGAE network

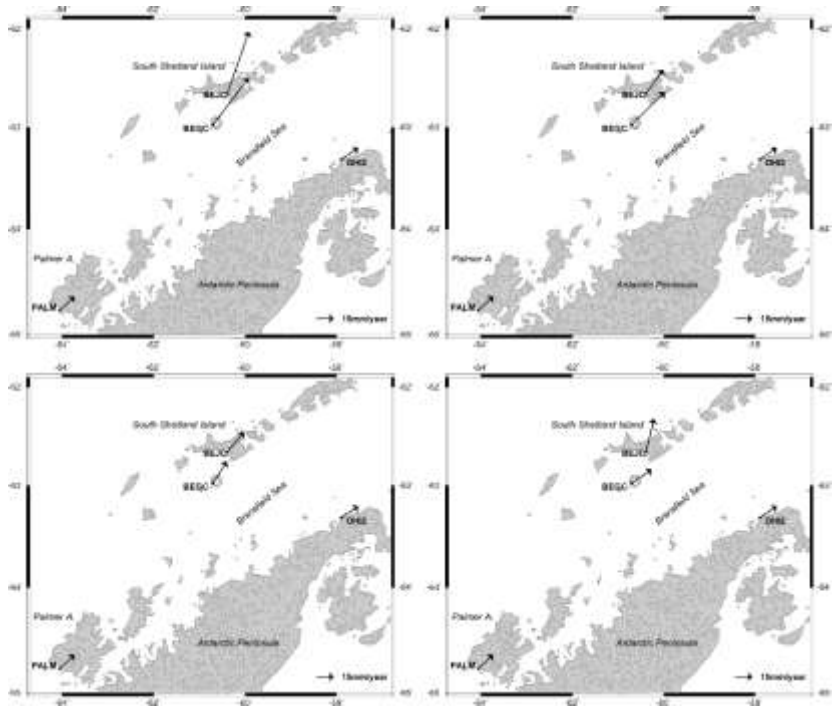

**Fig. 4** (from *left* to *right* and up to *bottom*) Tectonic deformation models for the periods 1999–2000 to 2001–2002, 2001–2002 to 2002–2003, 2002–2003 to 2003–2004 and 2003–2004 to 2004–2005

## 2 Geodetic Activities on Deception Island

The main goals of the geodetic research on Deception Island are the establishment of a local reference frame to get an accurate position of scientific data, the determination of a deformation model for the island and the detection and monitoring of the volcanic activity. In order to achieve these objectives a geodetic network and a levelling network were designed and established along with a local geoid.

The active volcanism in this area is mainly located in the expansion core of the Bransfield Rift and is characterized by the presence of some emergent volcanoes like Deception, Penguin and Bridgeman. In addition there are also numerous submarines volcanic which show different evolution stages, from the volcanic cone to its obliteration, generated by the existing tectonic of divergence (Canals et al. 1997b). These volcanic buildings are 15 km. base diameter and about 450 m. high. In addition, there are around 34 small buildings with a mean diameter of 2.5 km. These buildings are aligned following the main direction of the basin. The continuous volcanic activity in the area is manifested by historical eruptions in Deception and Penguin islands, together with the Bransfield Central Basin rifting and the magmatic processes in the rift during the last two million years. This fact suggests that the volcanic activity in the Bransfield basin goes on in continuous development (González-Ferrán 1991).

Deception Island is located in the beginning of the expansion axis of the Central Bransfield Basin. It is a horseshoe shape stratovolcano of 30 km. diameter in its submerged part and around 15 km of diameter in the emerged zone. The island has a maximum height of 1.5 km over the marine bottom, being its highest point Pond Mount, over 540 m on the sea level (Smellie 2001). It encloses an inner bay, Port Foster, which is a central flooded depression. Throughout its history the island has suffered several periods of different volcanic activity. At present it is the most active volcano of the South Shetlands Islands and the Antarctic Peninsula, with dated eruptions in 1848, 1967, 1969, 1912, 1917 and 1970 as it can be see in Fig. 5 (Mart´ı et al. 1996).

During 1967–1970 eruptive period, the high volcanic activity caused the destruction of the Chilean and British Scientific Bases, located in Pendulum Cove and Whaler's Bay respectively. These eruptions changed the morphology of the island, forming an islet that joined lately to the island around Telephone Bay (Fig. 6). In this period a great amount of ashes was emitted and deposited on the neighbouring islands, as it can be observed in Johnson glacier at Livingston Island. Due to this eruptive event and their destructive consequences the scientific activities by Argentinean, Chilean and British investigators were interrupted.

The activities to control and monitor the volcanic activity by means of geodesic and geophysical techniques were continued in 1986, when the first Spanish Antarctic campaign took place. Nowadays, the superficial evidences of volcanic activity are mainly the presence of fumarolic areas with 100°C and 70°C gaseous emissions in Fumarolas and Whaler's Bay respectively, 100°C hot soils in Cerro Caliente, and 45°C and 65°C thermal springs in Pendulum Cove and Whaler's Bay. In addition, there are numerous areas where a significant seismic activity is detected.

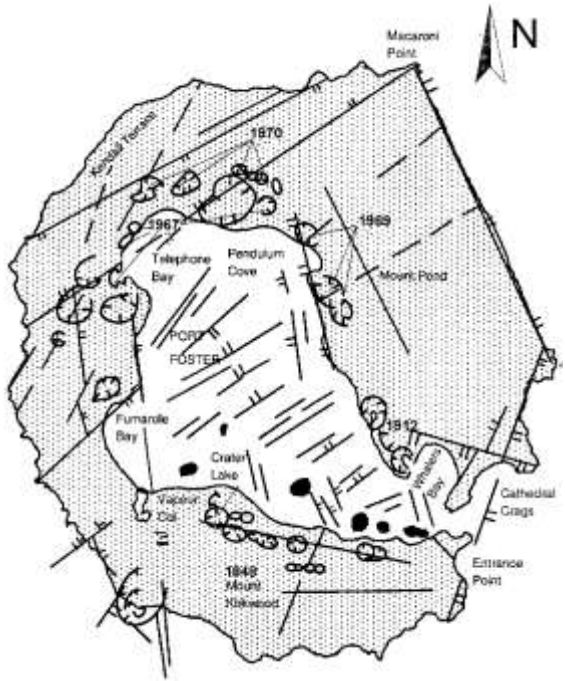

**Fig. 5** Deception Island tectonic setting with dated eruptions

This registered seismicity has two different sources: on one hand, the origin of tectonic activity due to the expansion of Bransfield Rift, and on the other hand, the purely volcanic source. In 1991–1992 and 1998–1999 campaigns two seismic-volcanic crises were detected (Ibáñez et al. 2003).

## 2.1 Regid Geodetic Network

The REGID geodetic network was established from GPS observations in order to set up the basic reference frame of both the observations and the results obtained by other sciences. In this way, the network has been the fundamental reference for the Spanish cartography of the Antarctica, for the geophysical and oceanographic observations and for the hydrographical sounding carried out throughout the Spanish campaigns at the Antarctica. On the other hand, GPS accuracy, especially for horizontal positioning, has made the REGID network very useful for the control of the geodynamic activity of the island.

The construction of the geodetic network began in the 1989–1990 Antarctic campaign. The aim of the studies and the place where they were going to be carried out led to the construction of high stability benchmarks, with no visual impact. They were built with concrete, well-rooted in the permafrost with steel bars and a low height above the ground. A standard screw was fixed in one of the corners of the post to allow the use of geodetic instrumentation. Furthermore, some of the stations were built using existing structures of the buildings destroyed by the volcanic activity which happened between 1967 and 1970 (Berrocoso 1997).

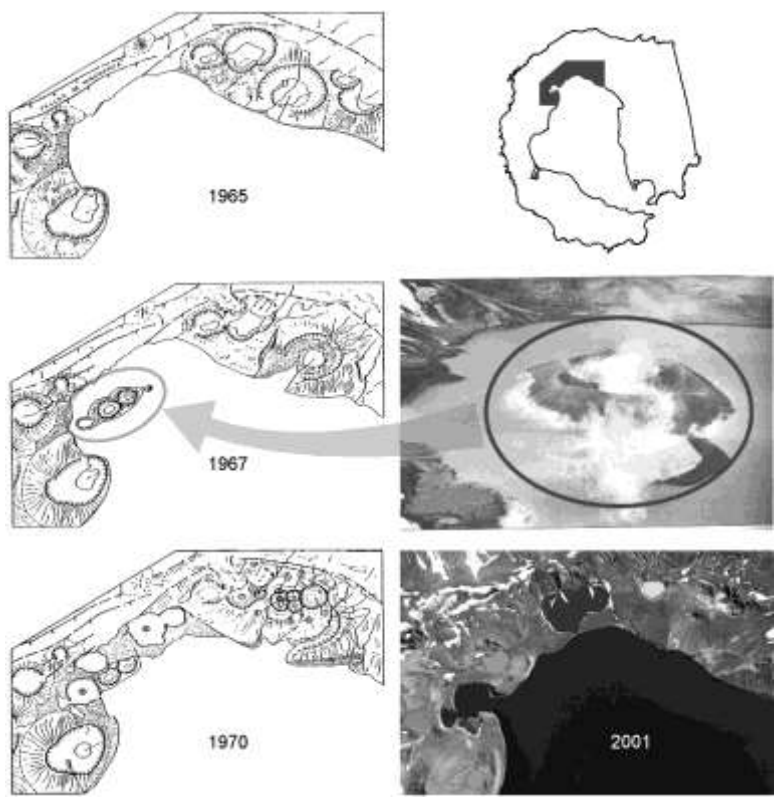

**Fig. 6** Morphological evolution in Telephone Bay during the 1967–1970 eruptive process (Brecher 1975)

In the 1990–1991 campaign, the geodetic network consisted of five stations, four of them in Deception Island and another one at the Spanish Base Juan Carlos I in Livingston Island. Their placements was chosen according to several factors, including accessibility, stability of the building, proximity of fumarolic areas, density of founded epicentres, last eruptions, etc. The stations in Deception Island were placed at the Argentinean Base (BARG) next to the Radio hut, at an existing hydrographical station in Fumaroles Bay (FUMA), at Pendulum Cove (PEND) near the remains of the Chilean Base and at the foundations of the old British Base in Whalers' Bay (BALL). BARG station was the main vertex of the network. The station in Livingston Island was built at the Spanish Antarctic Base Juan Carlos I.

During the 1995–1996 campaign a new station was built near the Spanish Antarctic Base Gabriel de Castilla. This station (BEGC) was set as the new main point of the network because it was further from one of the most active areas of the island, Fumaroles Bay. This station has an electricity supply and was built in the same way as the others (Berrocoso 1997). During the 2001–2002 campaign seven more stations were built in the north and south area of Deception Island, in order to fill in its entire inner ring and to make the REGID network more consistent. The designed of the stations is given in Fig. 4. The construction of these new stations was carried out according to the same geodetic directions as the last ones: clear GPS horizon, low multipath effect, etc. The area where they were built was chosen taking into account the volcanic nature of the island, but without considering any prior volcanic activity (Vila et al. 1992; García et al. 1997; Ibáñez et al. 2003).

These new stations are: UCA1, placed in northwest of the island, in the east of Obsidian Hill; TELE, located in the area of flooded craters in Telephone Bay; BOMB, positioned in a volcanic bombs field between 1970 Craters and Telephone Bay; CR70, placed in the area of craters generated by the eruptions in 1970 that destroyed the Chilean Base in Pendulum Cove; GLAN, located near the south of Black Glacier; GEOD, situated next to the Soto crater between Colatinas and the Spanish Antarctic Base Gabriel de Castilla; and COLA, sited in Colatinas, in the southeast part of the island. During the 2006–2007 campaign two more stations were included in the network, in Punta Collins, in the South of the island, and the levelling benchmark LN000, which has become a geodetic mark from now on. The final distribution of the network is shown in Fig. 7.

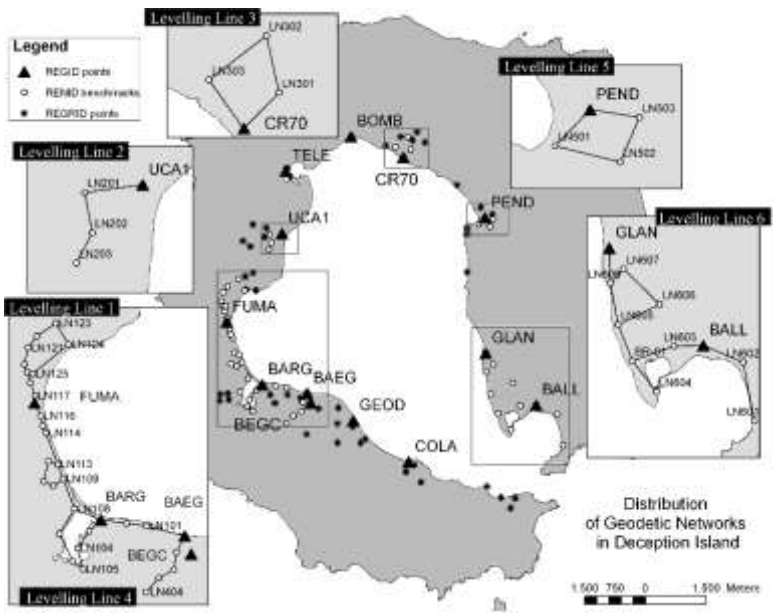

**Fig. 7** Distribution of Geodetic Networks in Deception Island

Stations of the initial network -BARG, BEJC, PEND, FUMA and BALL- have been observed during the Spanish Antarctic campaigns from 1991 to 1992. BEGC station was included during the 1995–1996 campaign, when it was observed for the first time. After the monumentation of seven new stations in the 2001–2002 campaign, COLA, GEODEC, UCA1, TELE BOMB, CR70 and GLAN, the whole network was observed during the austral summers from 2001 to 2002 campaign. The network has been surveyed until the last Antarctic campaign. During the 1991–1992 campaign, the fixed station was the one at the Argentinean Base (BARG). Observations from this station and from the one at Livingston Island (BEJC) were sent to the SCAR to be included in the SCAR92 solution.

In the 1999–2000 campaign, the five stations of the whole initial network were observed simultaneously for the first time. From 2002 to 2003 campaign, the joint BEGC-FUMA-PEND was maintained and the rest of the network was observed too.

The processing of the data has been made following the same methodology that previously with network RGAE with Bernese v4.2 software (Hugentobler et al. 2001). The IGS station OHIG was used to the adjustment of the network. Data from OHIG station are referred to the ones corresponding to the first of January of the ITRF96, ITRF97 and ITRF2000 systems, respectively. Every campaign was processed independently, fixing the coordinates of one of the stations of the network. BARG was the fixed station until 2001, and then it was changed to BEGC station.

Once the coordinates of the fixed station were set, the coordinates for the rest of the stations in the network were obtained by means of radial baselines between the fixed stations and the others. For the processing of the data, simultaneous observations were considered. To estimate the tropospheric delays we use the default meteorological data, since the only available data were the medium temperature and pressure and their inclusion could produce more errors than the models provided by the software. In the Table 3 results are referred to the ITRF2000, epoch 2003.1.

## 2.2 RENID Levelling Network

During the 2001–2002 Antarctic campaign a levelling network was designed to establish a reference frame for the real time monitoring of the vertical deformation taking place in the island. The network consists of six independent lines around the island, which allow a quick action just in the area where the volcanic activity has been detected. The distribution is shown in the Fig. 7 and in Table 4. The first bench mark to be monumented was the LN000 benchmark which to replace the BARG geodetic like origin of altitudes.

Levelling line 1 links the Spanish Base Gabriel de Castilla, the Argentinean Base Deception and Fumaroles Bay. It is supported by the LN000 bench mark and the geodetic stations BEGC; BARG and FUMA; levelling line 2, which extends along the Obsidian Hill, is supported in the geodetic station UCA1; levelling line 3 is supported by the geodetic station CR70 and it is placed in the area where the last eruption in 1970 took place; levelling line 4 is in the surroundings of the River Mekong and it is also supported on the main bench mark LN000; levelling line 5 is supported by PEND, in Pendulum Cove, and finally, levelling line 6 is supported by the geodetic stations GLAN and BALL, in Whalers Bay (Fig. 7).

The realization of the geometric levelling with sub-centimetre precision allows the establishments of a precise reference frame to measure instantaneous vertical deformation. Every bench mark in every line was positioned using a differential GPS. Table 4, shows the vertical data for the RENID levelling network. The LN000 benchmark is the reference point for levelling lines 1 and 4, and the geodetic stations UCA1, CR70, PEND and BALL are the reference points for lines 2, 3, 5 and 6, respectively.

The connections among lines were done during the 2002–2003 campaign, using a Wild T2000 theodolite and a DI5000 distancemeter. Lines 1, 2, 3 and 5 were connected among them, and levelling line 6 was linked to the geodetic station COLA, also connected to the LN000 bench in line 1 along the coastline. The connection between line 4 and line 1 was done during the last stage of that campaign.

## 2.3 REGRID Gravimetric Network

REGRID gravimetric network was established during the 2002–2003 Antarctic campaign from the geodetic stations of the REGID network and the levelling benchmarks of the RENID network. The fundamental gravimetric point in the island was set up to be GBEGC, in the surroundings of the Spanish Antarctic Base Gabriel de

**Table 3** REGID coordinates referred to ITRF2000 at 2003.1

| Station | X (mts.) | $\sigma_X$ | Y (mts.) | $\sigma_Y$ | Z (mts.) | $\sigma_Z$ |
| --- | --- | --- | --- | --- | --- | --- |
| BEGC | 1423027.714 | 0.001 | −2533143.956 | 0.001 | −5658977.759 | 0.001 |
| BARG | 1422140.963 | 0.001 | −2534034.247 | 0.003 | −5658736.141 | 0.003 |
| BALL | 1427933.006 | 0.003 | −2530514.413 | 0.003 | −5658856.008 | 0.006 |
| FUMA | 1421994.742 | 0.001 | −2535674.049 | 0.003 | −5658043.664 | 0.003 |
| BEJC | 1451089.543 | 0.003 | −2553226.299 | 0.003 | −5642854.597 | 0.006 |
| PEND | 1428697.709 | 0.001 | −2534792.701 | 0.003 | −5656765.170 | 0.003 |
| COLA | 1424578.900 | 0.003 | −2530853.317 | 0.003 | −5659569.913 | 0.009 |
| GLAN | 1427356.936 | 0.003 | −2532112.816 | 0.006 | −5658291.897 | 0.009 |
| GEOD | 1423777.232 | 0.003 | −2532298.366 | 0.003 | −5659121.635 | 0.009 |
| UCA1 | 1424097.820 | 0.003 | −2536792.788 | 0.003 | −5657026.513 | 0.009 |
| CR70 | 1427518.181 | 0.003 | −2536915.250 | 0.006 | −5656109.896 | 0.012 |
| TELE | 1424808.476 | 0.003 | −2537985.468 | 0.006 | −5656311.865 | 0.012 |
| BOMB | 1426582.129 | 0.003 | −2537924.128 | 0.006 | −5655895.118 | 0.015 |

**Table 4** Description of the RENID levelling network

| Levelling line | Levelling benchmark | Level difference | Levelling line | Levelling benchmark | Level difference |
|---|---|---|---|---|---|
| | LN101 | +2.132 | LINE 2 REF: | LN201 | +14.307 |
| | LN102 | +8.601 | | LN202 | +2.998 |
| | BARG | -2.868 | UCA1; | LN203 | +8.968 |
| | LN103 | -0.739 | LINE 3 REF: | LN301 | +7.283 |
| | LN104 | -3.464 | | LN302 | +21.333 |
| | LN105 | +4.971 | CR70 | LN303 | +1.437 |
| | LN106 | +1.177 | LINE 4 REF: | LN401 | +4.642 |
| | LN107 | +2.861 | LN000 | LN402 | +12.430 |
| | LN108 | +0.289 | | LN403 | +20.676 |
| | LN109 | +3.797 | | LN404 | +32.498 |
| | LN110 | +20.390 | LINE 5 REF: | LN501 | -5.998 |
| | LN111 | +24.307 | FUMA | LN502 | +15.621 |
| LINE 1 REF: | LN112 | +10.304 | | LN503 | +22.493 |
| LN000 | LN113 | -1.745 | | BR-01 | +16.380 |
| | LN114 | -4.264 | | LN601 | -5.113 |
| | LN115 | -4.917 | | LN602 | +1.039 |
| | LN116 | -4.066 | | LN603 | +7.947 |
| | FUMA | -2.468 | LINE 6 REF: | LN604 | -4.610 |
| | LN117 | -4.010 | BALL | LN605 | +8.684 |
| | LN118 | -2.352 | | LN606 | +11.331 |
| | LN119 | -2.201 | | LN607 | +17.915 |
| | LN120 | -0.642 | | LN608 | +7.061 |
| | LN121 | +2.688 | | GLAN | +7.161 |
| | LN122 | +5.515 | | | |
| | LN123 | +11.757 | – | | |
| | LN124 | -0.220 | | | |
| | LN125 | -4.392 | | | |

Castilla. Another gravimetric point at Livingston Island, GBEJC, is included in the network.

Gravimetric measurements were obtained with a Lacoste & Romberg D-203 gravimeter with a priori deviation of 10 nms$^{-2}$. A gravimetric link among GBEGC (fundamental point) in Deception Island, GBEJC in Livingston Island and APPA gravimetric point in Punta Arenas (Chile) was made, which value is 9813208100 nms$^{-1}$. Tides, height and drift corrections were applied to the whole set of gravimetric data. The distribution of the secondary points is given in Fig. 7. The values for gravimetric measurements for this network is given in Table 5.

**Table 5** Gravimetric measurements for the REGID network geodetic stations

| Station | g [nm/s$^{-2}$] | σ[nm/s$^2$] |
|---|---|---|
| BALL | 9822071743 | 138 |
| BARG | 9822106750 | 174 |
| BEGC | 9821962801 | 145 |
| BOMB | 9822075418 | 259 |
| COLA | 9822089209 | 175 |
| CR70 | 9822031402 | 259 |
| FUMA | 9822079377 | 108 |
| GBEGC | 9822044682 | 20 |
| GEOD | 9822061136 | 117 |
| GLAN | 9822028138 | 198 |
| PEND | 9822040199 | 160 |
| TELE | 9822077750 | 198 |
| UCA1 | 9822035897 | 198 |

## *2.4 Geodetic Network REGID Aplication for Volcanic Deformation Model Determination and Sources Location Estimation*

Superficial deformation in volcanic areas is interpreted like reflection of the changes of pressure in the superficial magmatic cameras. Greater deformations happen in limited areas and its interpretation, in terms of schematic models, implies to assume the existence of those superficial or brief coves or reservoirs, intrusions or forts increases of pressure without justification. A possible hypothesis indicates that the observed superficial displacements are the answer of a half elastic space to a pressure increase in magmatic cove. In order to determine the most probable models different geometries and different restrictions are considered (Lavalle et al. 2004; Abidin et al. 1998; Bock et al. 1997).

Deformation models representing Deception Island volcanic activity have been obtained from the GPS observations of the stations conforming the geodetic network REGID. GPS observations have been episodically made from 1991–1992 campaign to present. Taking into account that GPS system provides twice the accuracy in horizontal positioning than in the vertical component, we distinguish between the horizontal and the vertical models, although it is essential to conjugate both results for a proper interpretation. (Donnellan 1993; El-Fiky et al. 1999; Calais et al. 2000, Mantovani et al. 2001; Murray and Wooller 2002).

To obtain the deformation models a topocentric system was established with origin at BARG. In order to obtain the absolute deformation of the island BEJC station on Livingston Island was considered. Data from GPS surveying in Deception Island were processed using the Bernese v4.2 software according the methodology described on previous paragraphs, this is, precise orbits were used downloaded from several sources: SIO (*Scripps Institution of Oceanography*) for 1991–1992 campaign and CODE (*Centre of Orbits Determination*) for the rest of campaigns.

Every campaign was processed independently, fixing the coordinates of one of the stations of the network. BARG was the fixed station until 2001, and then it was changed to BEGC station. In the 1991–1992 campaign, the coordinates for the fixed station BARG were obtained from the international campaign SCAR92. Its precise absolute coordinates for the station were referred to the ITRF91, epoch 2.2.

To get these coordinates for the next campaigns, in 1995–1996 and 1999–2000, the coordinates of the IGS station OHIG corresponding to ITRF96, epoch 96.1 and to ITRF97, epoch 99.9, were fixed. It is situated at the Chilean Base O'Higgins in the Antarctic Peninsula, 150 km. away from Deception Island.

Due to the proximity of the BEGC station to the Spanish Base Gabriel de Castilla, the receiver is easily maintained and it is possible to collect data continuously during the campaign. That is why this station was considered as the main one in the network REGID from the 2001–2002 and 2002–2003 campaigns. Its coordinates were obtained by fixing the ones of OHIG corresponding to the ITRF2000, epoch 2002.1 and 2003.1. Once the coordinates of the fixed station were set, the coordinates for the rest of the stations in the network were obtained by means of radial baselines between the fixed stations and the others. For the processing of the data, simultaneous observations were considered.

In the 1991–1992 campaign, the radial processed baselines were BARG- FUMA, BARG- PEND and BARG- BALL. The ambiguities were calculated by the SIGMA strategy of the BERNESE software, since the observations were made by single-frequency receivers. Finally, the solution is referred to the ITRF91; in 1995–1996, observations of both L1 and L2 frequencies were available, so the ambiguities resolution was solved applying the QIF strategy. The final solution was referred to the ITRF96; in the 1999–2000, the baselines were processed altogether since it was the first time that every station was observed simultaneously. The ambiguities were also resolved with the QIF strategy and the final solution was referred to the ITRF97; from 2001–2002 and 2002–2003 campaigns, solutions for every week were combined with the ADDNEQ program of the BERNESE software. QIF strategy was also used and the final solution was referred to the ITRF2000.

Displacements models are calculated comparing absolute topocentric coordinates, obtained on every observation epoch, between two consecutives campaigns (Fig. 8); determining the velocity fields and its gradient by a finite elements interpolation. Horizontal deformation tensors are calculated to obtain deformation parameters, extension-compression area (Fig 9); finally the components in directions NS and EW are obtained (Dermanis 1985; Eren 1984; Van´ıcek and Krakiwsky 1986).

Mathematical horizontal deformation models are obtained from these displacements models by dilatation. (Bibby 1982; Drew y Snay 1989; Grafarend and Voosoghi 2003; Dzurisin 2006). From the analysis of the 1991–1992 to 2002–2003 campaign data we obtained a radial extensional process from 1995/1996 to 1999/2000 campaign. Afterwards, the process became compressive; there was an uplift process from 1995/1996 to 2001/2002 campaigns; and subsidence process before and after this period. Extensional process follows the NW-SE direction (Fracture Hero Zone direction), while the compressive process follows the Bransfield Rift extensional direction (Fig. 10).

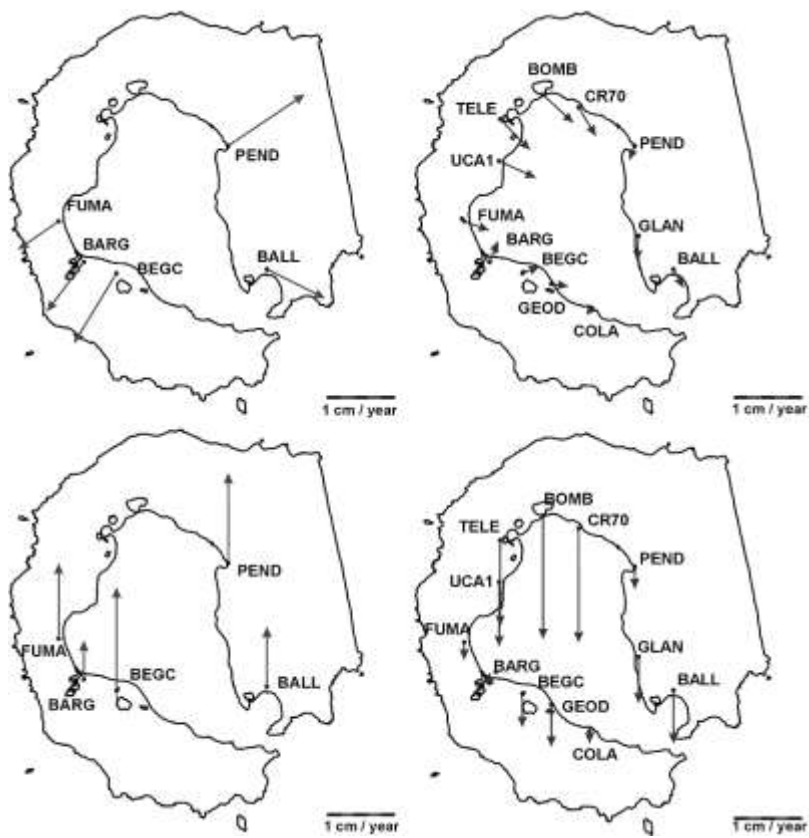

**Fig. 8** REGID geodetic network, RENID levelling network and REGRID gravimetric network distribution

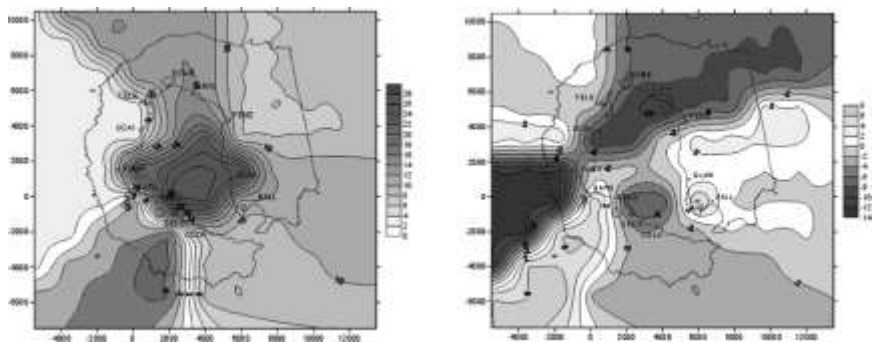

**Fig. 9** From *left* to *right, top* to *bottom*: horizontal and vertical deformation models for 1996.2/1999.9 and 2002.2/2003.2 epochs (cm./year), respectively

Changes detected from volcanic deformation models in extension-compression as well as subsidence-uplift-subsidence and changes in activity axes direction overlap with 1998/1999 seismovolcanic crisis.

Considering the initial configuration of the geodetic network REGID, Mogi model has been considered for the determination of the source location and variation for the hydrostatic pressure explaining the superficial deformation (Mogi 1958). Experimental data, horizontal and vertical displacements were used, whose Lame elasticity modulus and Poisson coefficient were fixed from theoretical data (Willson 1980; Newhall and Selft 1982; Bonaccorso 1996). Simulated Annealing inversion algorithm has been used for source estimation (Cervilli et al. 2001; Sambridge and Mosegaard 2002).

From 1992 to 1996, displacements can be due to the activity of principal source located in the interior ring, BARG-FUMA-PEND triangle (Fig. 11). From 1992 to 1999, a northern source is estimated at FUMA-PEND direction. The seismovolcanic crisis in 1998 seems to be produced by a chamber activity located at north of Port Foster, between Fumarolas Bay and Pendulum Cove. From 1999, the source is located deeper, indicating volcanic activity decreased, at least, until 2003 (Fernández-Ros 2006).

## 2.5 Determination of a Local Experimental Geoid

The presence of active volcanism in Deception Island and the existence of both superficial and deep seismicity make the tectonic situation of the area become

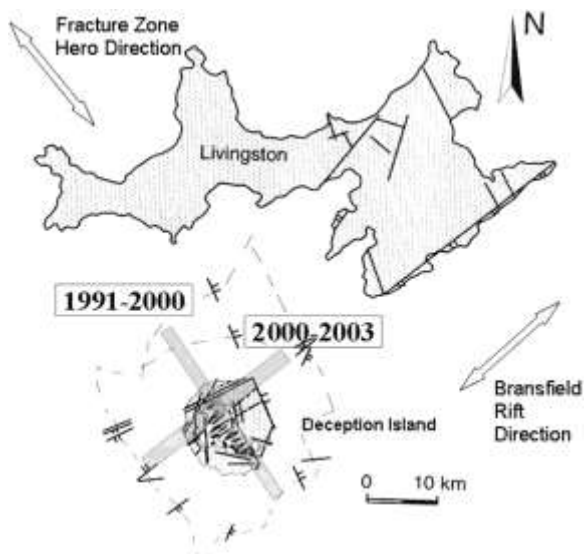

**Fig. 10** Superficial dilatation (ppm/year) from 1996.2/1999.9 and 2002.2/2003.2 epochs

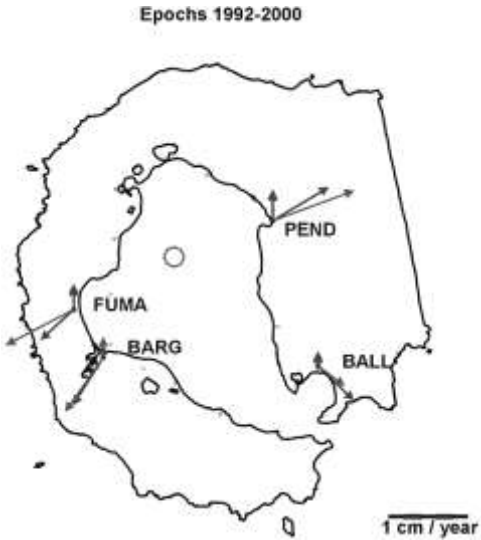

**Fig. 11** Volcanic activity responds to two main alignments for Deception Island: Source location estimated for superficial deformation origin zone from volcanic activity in Deception Island

very complex and, therefore, the main goal of the Geosciences studies. Thus, the establishment of a proper geodetic reference frame acquires a high degree of relevance.

The lack of physical meaning of the ellipsoidal height as well as the lack of accuracy in vertical deformation models make not only a mathematical but also a physical reference frame, such us the geoid, necessary in order to calculate the existing deformation. However, global geoid models, such as OSU91A or EGM96, are so inaccurate at small spatial scales in volcanic areas with high geodynamic activity, that determining an experimental geoid model to refine the globally extrapolated models is essential.

This experimental geoid will allow us to detect gravity field changes due to crustal movements and it will also make the determination of a hazard map possible thanks to the combination of gravity equipotential surfaces and data from a digital elevation model. In addition to this, the geoid could permit us to carry out real time levelling measurements by means of GPS receivers. In the area of Deception Island this application would be very interesting to measure the vertical deformation in case of volcanic reactivation in a fast and accurate way, which is almost impossible and dangerous using classical levelling methods.

Therefore, an experimental geoid model for Deception Island, computed from GPS data, geometrical levelling and absolute gravimetric measurements, has been calculated. This mean sea level surface will be a very important reference frame to enhance our knowledge of the area and will allow us to improve the accuracy of the high degree geopotential models in this area of the Antarctic.

Besides the main purpose of each one of the designed networks at Deception Island, other different and complementary measurements were gathered in order to use all the stations and benchmarks for any geodetic application, as the geoid determination. So, the stations of the REGID network were linked to the RENID network by geometric levelling, in such a way that these stations are now provided with levelling measurements with respect to the main benchmark LN000. In the same way, every levelling benchmark was positioned in 'fast-static' mode, keeping one base receiver at BEGC station. Therefore, and taking into account that the gravimetric network was established from REGID and RENID networks, REGRID is provided not only with absolute gravity values, but also with geodetic coordinates and levelling measurements.

Since REGID, RENID and REGRID networks had to be developed around the inner bay and not further than one kilometre offshore, 44 secondary points were also established to spread out the observations to the outer area and to make the measurement set denser. So, GPS observations, levelling and gravimetric measurements were also obtained for these marks; on one hand, every point was positioned by 'fast-static' mode. Regarding GPS data, although the REGID geodetic network has been surveyed from 1988, considered data correspond to the 2002–2003 Antarctic campaign, from November 2002 to February 2003 (Berrocoso et al. 2004a,b).

The adjustment of the network was carried out in two stages: firstly, absolute geocentric coordinates for both BEGC and BEJC stations were obtained by the processing of the network with the IGS station OHI2 respect to ITRF2000 reference frame, epoch 2000; secondly, the coordinates of the remaining stations at Deception Island, calculated respect to ITRF2000 as well, were processed from the coordinates previously obtained.

In relation to levelling measurements, levelling surveys were carried out using a Leica NA2 level, with $\sigma = 0.07$ mm, during the 2001–2002 and the 2002–2003 Antarctic campaigns. Corrections due to refraction effects were applied to the data whereas the sphericity effect was not taken into account because of the size of the area under study. Three different methods were considered to obtain this physical reference frame: remove-restore, collocation and GPS/levelling methods. GPS and gravimetric data are needed when using the two first methods whereas levelling data are also needed with the last one. The problem is that these data are really difficult and expensive to obtain in areas like the Antarctica.

On one hand, the remove-restore method is based on the resolution of the Stokes integral by breaking down the frequency spectrum of the gravity anomaly and undulation into different bands. Then, FFT or wavelets can be used as an integration method. On the other hand, collocation method provides values of the geoid height which best fit the gravity measurements by means of the matrix of covariance. Finally, GPS/levelling technique considers a set of points around the island where absolute gravity, levelling and ellipsoid coordinates are known. By means of the orthometric height in BARG, we are able to obtain the orthometric heights in the remaining points. So, to calculate the geoid, we only have to consider the well-known

formula h = H + N, where h, H and N are the ellipsoidal, orthometric and geoid heights respectivel

With regards to computational cost, remove-restore method has a higher cost than GPS/ levelling and collocation methods not only due to the use of Global Geopotential Models but also because of the methods of numerical integration. Also, by comparing the acquired values to those geoid heights from global geopotential models, we easily obtain the correction factor for the area of Deception Island. In fact, GPS/levelling and collocation techniques provide more details than remove-restore method in comparison with global geopotential models (Fig. 12).

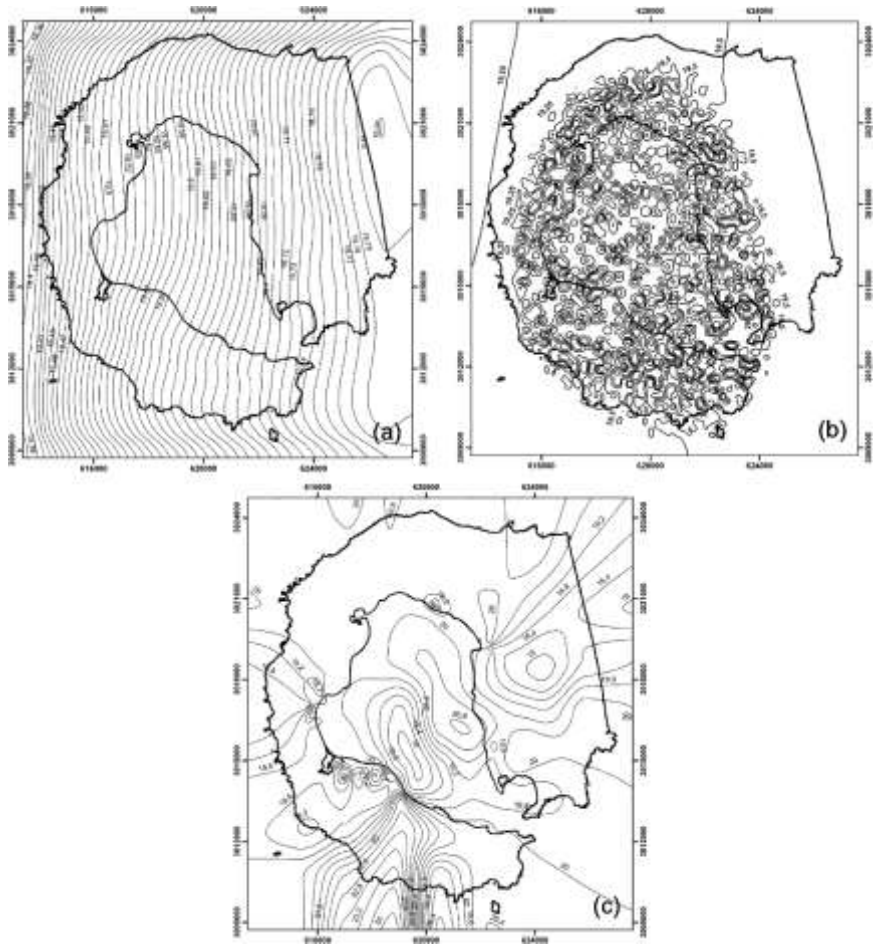

**Fig. 12** Geoid height maps: (**a**) remove-restore; (**b**) collocation; (**c**) GPS/levelling

## 3 Summary and Conclusions

The geodetic networks which have been established in South Shetland Islands, Bransfield Strait and the Antarctic Peninsula have been presented and described in this work. The main aims of RGAE and REGID networks are, on the one hand, to define a proper geodetic reference frame in the area under study and, on the other hand, to make possible the survey of tectonic and volcanic processes taking place in the region. We would like to point out the fact that the OHI2 and PALM stations are not the same as those ones belonging to the current IGS network although they are near the present points. In order to keep the Spanish base located in the island and the deformation of the volcano under surveillance, an accurate levelling network (RENID network) has been also designed and built. These benchmarks will allow us to monitor the real time island deformation as a volcanic precursor.

So as to define a physical reference frame for Deception Island, an experimental geoid model, computed from GPS data, geometrical levelling and absolute gravimetric measurements, has been calculated. This mean sea level surface will be necessary to enhance our knowledge of the area and will allow us to improve the accuracy of the high degree geopotential models in this zone of the Antarctic.

Finally, the need to establish an Antarctic geoid with decimetric accuracy, the need to homogenize the existing cartography, to determine local and regional refraction models in the Antarctica and the need to make the networks in existence denser, they all define the objectives with top priority in the geodetic Antarctic research which must be favoured from the International Polar Year.

**Acknowledgments** The realization of this geodetic research is supported by the Spanish Ministry of Education and Science as part of the National Antarctic Program. The following research projects have been awarded so far: "Recognition and fast evaluation of the volcanic activity of the island Deception (GEODESY) (ANT1999.1430.E/HESP)"; "Geodetic Studies on Deception Island: deformation models, geoid determination and Information System for Scientific (REN2000.0551.C03.01/ANT)"; "Acquisition of a scientific software for GPS data processing (REN2000.2690.E)"; "Geodetic Control of the volcanic activity of Decepcion Island (CGL2004.21547.E/ANT)"; "Update of the Spanish Cartography for Decepcion Island (CGL2004.20408.E/ANT)"; "Volcanotectonic activity on Deception Island: geodetic, geophysical investigations and Remote Sensing on Deception Island and its surroundings (CGLl2005-07589-c03-01/ANT)"; "Geodetic Control of the volcanic activity of the Island Deception". We would like also thank the BIO Hespérides and the Palmas crew for they support in the development of the carried out activities, the logistic crew of both Spanish Antarctic Bases Gabriel de Castilla and Juan Carlos I., the Hydrographic Institute of the Navy, the Geographic Service Army as well as the Royal Naval Observatory of San Fernando.